# Effect of entropy on the dynamics of supercooled liquids: New results from high pressure data

R. CASALINI†‡ and C.M. ROLAND†

†Naval Research Laboratory, Chemistry Division, Washington, DC 20375-5342
‡George Mason University, Chemistry Department, Fairfax, VA 22030

(July 18, 2006)

Abstract

We show that for arbitrary thermodynamic conditions, master curves of the entropy are obtained by expressing *S(T,V)* as a function of $TV^{\gamma_G}$, where *T* is temperature, *V* specific volume, and $\gamma_G$ the thermodynamic Grüneisen parameter. A similar scaling is known for structural relaxation times, $\tau = \Im\left(TV^{\gamma}\right)$; however, we find $\gamma_G < \gamma$. We show herein that this inequality reflects contributions to *S(T,V)* from processes, such as vibrations and secondary relaxations, that do not directly influence the supercooled dynamics. An approximate method is proposed to remove these contributions, $S_0$, yielding the relationship $\tau = \Im_1\left(S - S_0\right)$.

## 1. Introduction

If cooled fast enough to avoid the crystallization, any liquid can assume a metastable "supercooled" state. Further cooling causes progressive increases of the structural relaxation time τ and viscosity η of the liquid, until a temperature, the glass temperature, is reached in which the η is so large as to be considered a solid (i.e., a glass), while τ exceeds the typical experimental timescale. Analogously a glass can be formed by compressing under hydrostatic pressure a liquid while maintaining its temperature fixed. The study of the glass transition along these two pathways has proven fruitful for understanding the mechanisms underlying the glass transition, including resolving and quantifying the relative effect of $T$ and $V$ [1,2,3]. A key idea to progress in this area was the determination that at $T_g$ (or some other characteristic temperature for which τ is constant [4, 5, 6]), the product of temperature times specific volume, with the latter raised to a power, was constant; i.e., $T_g V_g^{\gamma} = constant$ with $\gamma$ a material-specific constant [7, 8]. Quite recently this has been generalized, with the discovery that the behavior of $\tau(T,V)$ throughout the supercooled regime can be described as [9]

$$\tau(T,V) = \Im\left(TV^{\gamma}\right) \tag{1}$$

where $\Im$ is a generic unknown function. This behavior has been verified by several groups using different techniques and finding comparable values of γ for the same material [6]. The original idea motivating this approach was the property of the inverse power law potential $U(r) \propto r^{-n}$ (with $r$ the particle distance), for which $\tau(T,V) = \tilde{\Im}\left(TV^{n/3}\right)$ [10]. Thus, for a 6-12 Lennard-Jones type potential, if the local properties are dominated by the repulsive part of the potential, eq.(1) is expected with γ=4. Such behavior has been verified for OTP [11]. For other materials γ assumes various values in the range $0.18 \leq \gamma \leq 8.5$ [6].

A still unresolved issue of the past half century is whether (and how) the dynamics of supercooled liquids and the consequent glass transition can be related to thermodynamic quantities, and even if there is an underlying thermodynamic transition [12]. The scaling law (eq.(1)) implies that if this process is dominated by a thermodynamic quantity, than the latter has to be a function of both $T$ and $V$, satisfying the same scaling law. Recently, we have shown that under the approximations that (i) the isochoric heat capacity of a liquid, $C_V$ is constant with respect to T and (ii) the difference between isobaric and isochoric heat capacities, $C_P$ - $C_V$, is constant with respect to $V$, than the entropy $S$, for a supercooled liquid over typical experimental ranges satisfies a similar scaling law [13]

$$S(T,V) = \widehat{\Im}\left(TV^{\gamma_G}\right) \tag{2}$$

where $\gamma_G$ is the thermodynamic Grüneisen parameter given by

$$\gamma_G = \frac{V\alpha_P}{C_V\kappa_T} \tag{3}$$

with $\alpha_P$ the isobaric expansion coefficient, $\kappa_T$ the isothermal compressibility, and $C_V$ the isochoric heat capacity. For the case of the inverse power law potential the entropy also scales as [14]

$$S \cong \Im_1\left(TV^{n/3}\right) \tag{4}$$

Therefore, in this particular case, $\gamma_G = \frac{n}{3}$, and $\tau$ should be a function of $S$ alone.

In the following, we used experimental data to calculate the entropy in different condition of $T$ and $V$ to confirm eq.(2) and the consequent relation between $\tau$ and $S$.

## 2. Method

$S$ at any condition of $T$ and $P$ can be calculated from thermodynamic data using

$$S(T,P) = S_{ref}(T_{ref}, P=0) + \int_{T_{ref}}^{T} \frac{C_P}{T} dT - \int_0^P \left.\frac{\partial V}{\partial T}\right|_P dP \tag{5}$$

where $S_{ref}$ is the entropy at a reference temperature $T_{ref}$ at atmospheric pressure (here $P$=0). According to the Tait equation of state (EOS) [15]

$$V(T,P) = V(T,P=0)\left\{1 - C\ln\left[1 + P/\left(b_1\exp\left(-b_2T\right)\right)\right]\right\} \tag{6}$$

where $C$=0.0894 and $b_1$ and $b_2$ are constants. The integral of the thermal expansivity in eq.(5) can be calculated as

$$\int_0^P \left.\frac{\partial V}{\partial T}\right|_P dP = P\left[\left.\frac{\partial V}{\partial T}\right|_{P=0}(1+C) - V(T,0)Cb_2\right] + CB(T)\ln\left(1+\frac{P}{B(T)}\right)\left[V(T,0)b_2 - \left(1+\frac{P}{B(T)}\right)\left.\frac{\partial V}{\partial T}\right|_{P=0}\right] \tag{7}$$

Therefore, after calculating $S$ at atmospheric pressure from $C_P$, we calculate $S$ at any $T$ and $P$ using eq.(7) with the Tait parameters. This procedure is very similar to that used previously to test the Adam Gibbs model [16, 17].

## 3. Results and Discussion

The insert to figure 1 shows the change of entropy $S$-$S_{ref}$ for salol versus the specific volume $V$, taking as $S_{ref}$ the entropy at $T$=220K and atmospheric pressure (for which τ~10s). The entropy at atmospheric pressure was calculated using the data from refs.[18] and [19]. To test eq.(2), we plotted $S$-$S_{ref}$ versus the product of $T$ and $V$ with $V$ raised to an exponent. There is a satisfactory collapse of the $S(T,V)$ data when the exponent is 1.7, as seen in figure 1. This empirical value of 1.7 is close to that previously determined for salol using eq.(3) with thermodynamic data , $\gamma_G$=1.9 [13].

A second example, polyvinylacetate (PVAc), is shown in figure 2, for which the entropy was calculated using the data from [20] and [21], $S_{ref}$ is the entropy at T=313K and atmospheric pressure (again τ~10s). These data also collapse onto a master curve versus $TV^{0.9}$. Using eq.(3), previously we found using eq.(3) that $\gamma_G \approx 0.7$ [13]; therefore, eq.(2) is verified to an acceptable degree of accuracy for PVAc.

Another case for which the data are not shown is propylene carbonate. Using the data of [22, 23], we found superpositioning of the $S(T,V)$ versus $TV^2$. Eq.(3) gives $\gamma_G$=1.4. [13].

Thus the behavior of $S(T,V)$ is described reasonably well as a function of $TV^{\gamma_G}$ (eq.(2)), with the value of $\gamma_G$ obtained from eq.(3). This supports the approximations used to arrive at these expressions [13]. However, the value of the exponent $\gamma_G$ yielding collapse of the $S(T,V)$ data is significantly smaller than the parameter $\gamma$ (eq.(1)) used to superpose relaxation times: $\gamma$=5.2 for salol [9], $\gamma$=2.5 for PVAc [24] and $\gamma$=3.7 for PC [22, 23]. This difference indicates that there is not a simple, direct connection between the relaxation times and the entropy changes accompanying vitrification. This is a well-known problem, arising from contributions to the entropy from other motions, such as vibrations and local secondary relaxations involving, for example, pendant moieties in polymers [25, 26, 27], which we refer to herein as $S_0$. The problem of subtracting $S_0$ is not straight forward, since these contributions in principle are both *T*- and *V*-dependent. However, the fact that the scaling exponent for $S(T,V)$ is roughly one-third of the value of $\gamma$ for $\tau(T,V)$, is an indication that $S_0$ has a relatively weaker dependence on *V*. This is consistent with the observation that the characteristic times of local secondary motions are less sensitive to *V* compared to that of structural relaxation (which is not true for the activation energies).[28] Therefore, as a first approximation we assume that as pressure varies, these extra contributions remain unchanged; that is, $S_0(T,V) \sim S_0(T)$. $S_0$ is taken as the value of $S(T_g)$ (corresponding to the dielectric $\alpha$-relaxation time =10s), for different *T* as a function of *V*. In figure 3 and 4 the respective dielectric relaxation times for salol [29] and PVAc [30, 31] are shown versus $S\text{-}S_{ref}$. In the inserts are plotted the values of $S\text{-}S_{ref}$ for which $\tau$=10s (i.e. $S_0\text{-}S_{ref}$) versus *T*; the relationship is linear. Subtracting $S_0\text{-}S_{ref}$ from $S\text{-}S_{ref}$, we obtain the portion of the total entropy, $S\text{-}S_{ref}$, associated only with structural relaxation. In figure 5 $S\text{-}S_{ref}$ (open symbols) and $S\text{-}S_0$ (solid symbols) are shown for salol and PVAc. The *T*-dependences are similar and interestingly $S\text{-}S_0$ is about one-half $S\text{-}S_{ref}$ for both materials.

In figure 6 the dielectric relaxation times from figures 3 and 4 are plotted versus $S\text{-}S_0$. For both salol and PVAc, the data essentially superpose to from a single curve. This demonstrates a direct correspondence between $S\text{-}S_0$ and $\tau$, and thus rationalizes the $TV^{\gamma}$–dependences of the dynamics and the entropy.

## 4. Conclusions

In the present paper we have investigated the behavior of $S(T,V)$ using literature data to determine if the entropy is a function of $TV^{\gamma_G}$ ((2)) with $\gamma_G$ given by eq.(3). Our purpose is to investigate whether the behavior is related to the scaling relation for $\tau(T,V)$ (i.e., eq.(1)). For three glass formers, salol, PVAc, and propylene carbonate, the entropy is found to vary uniquely with $TV^{\gamma_G}$, with values of the parameter $\gamma_G$ quite consistent with eq.(3). The differences between $\gamma_G$ and $\gamma$ are reconciled with the excess contribution to the total entropy from secondary processes that are not arrested at the glass transition and vibrations, which do not affect structural relaxation. We assume herein that this excess entropy, $S_0$, has a negligible dependence on $V$, which allows its direct determination from high pressure data. Specifically, we take the value of $S$ for which $\tau$ =10s at different $T$ and $V$. After subtracting $S_0$ from the total entropy, $\tau$ is found to depend directly on the remaining part of the entropy, $\tau = \Im_1\left(S - S_0\right)$, as shown in figure 6. This implies that $S$-$S_0$ scales with the same exponent $\gamma$ as does $\tau$. As shown in figure 7, $S$-$S_0$ exhibits a linear variation on $1/TV^{\gamma}$; moreover, there is no suggestion of a thermodynamic divergence in the behavior above absolute zero.

In conclusion a possible interpretation of the scaling of the relaxation times (eq.(1)) in terms of the entropy is investigated. We verify that the entropy conforms to the scaling relation of the relaxation times; however, the same scaling exponent is obtained only if that portion of the entropy arising from other dynamical processes is subtracted. We also demonstrate a method to determine $S_0$.

## 4. Acknowledgements

This work was supported by the Office of Naval Research. Inspiring conversations with S. Capaccioli and J.C. Dyre are gratefully acknowledged.

Figure captions

Figure 1. Entropy of salol at constant pressure and three constant temperatures minus $S_{ref}=S(T=220\text{K}, P=0.1\text{MPa})$ versus $TV^{1.7}$ and (insert) versus the specific volume. The entropy was calculated from $C_P$ at atmospheric pressure [18,18] and $V(T,P)$ data [18,19].

Figure 2. Entropy of polyvinylacetate at constant pressure and three (constant) temperatures minus $S_{ref}=S(T=313\text{K}, P=0.1\text{MPa})$ versus $TV^{0.9}$ and (insert) versus the specific volume. The entropy was calculated from $C_P$ at atmospheric pressure [21,22] and $V(T,P)$ data [20,21].

Figure 3. Dielectric relaxation times versus the entropy after subtraction of $S_{ref}$ (= $S(T=220\text{K}, P=0.1\text{MPa})$, at atmospheric pressure and three fixed temperatures ($\tau$ from [27,29]). In the insert are $S-S_{ref}$ at $\tau$=10s versus temperature as determined from the data in the main figure. The solid line is a linear fit.

Figure 4. Dielectric relaxation times versus the entropy after subtraction of $S_{ref}$ (=$S(T=313\text{K}, P=0.1\text{MPa})$, at atmospheric pressure and three fixed temperatures ($\tau$ from [28, 29,30,31]). In the insert are $S-S_{ref}$ at $\tau$=10s versus temperature as determined from the data in the main figure. The solid line is a linear fit.

Figure 5. Top: open symbols are T-dependence of entropy of salol at atmospheric pressure minus $S_{ref}=S(T=220\text{K}, P=0.1\text{MPa})$. Solid symbols are S minus $S_0 = S|_{\tau=10s}$ determined from the fits of the data in the insert of Figure 3.
Bottom: open symbols are T-dependence of entropy of PVAc at atmospheric pressure minus $S_{ref}=S(T=313\text{K}, P=0.1\text{MPa})$. Solid symbols are $S$ minus $S_0 = S|_{\tau=10s}$ determined from the fits of the data in the insert of Figure 4.

Figure 6. Dielectric relaxation times for salol and PVac (reported in figures 3 and 4, respectively) versus $S-S_0$.

Figure 7. $S-S_0$ versus $TV^{\gamma}$ with the value of $\gamma$ for each material indicated in the figure. Symbols are the same as in figure 6.

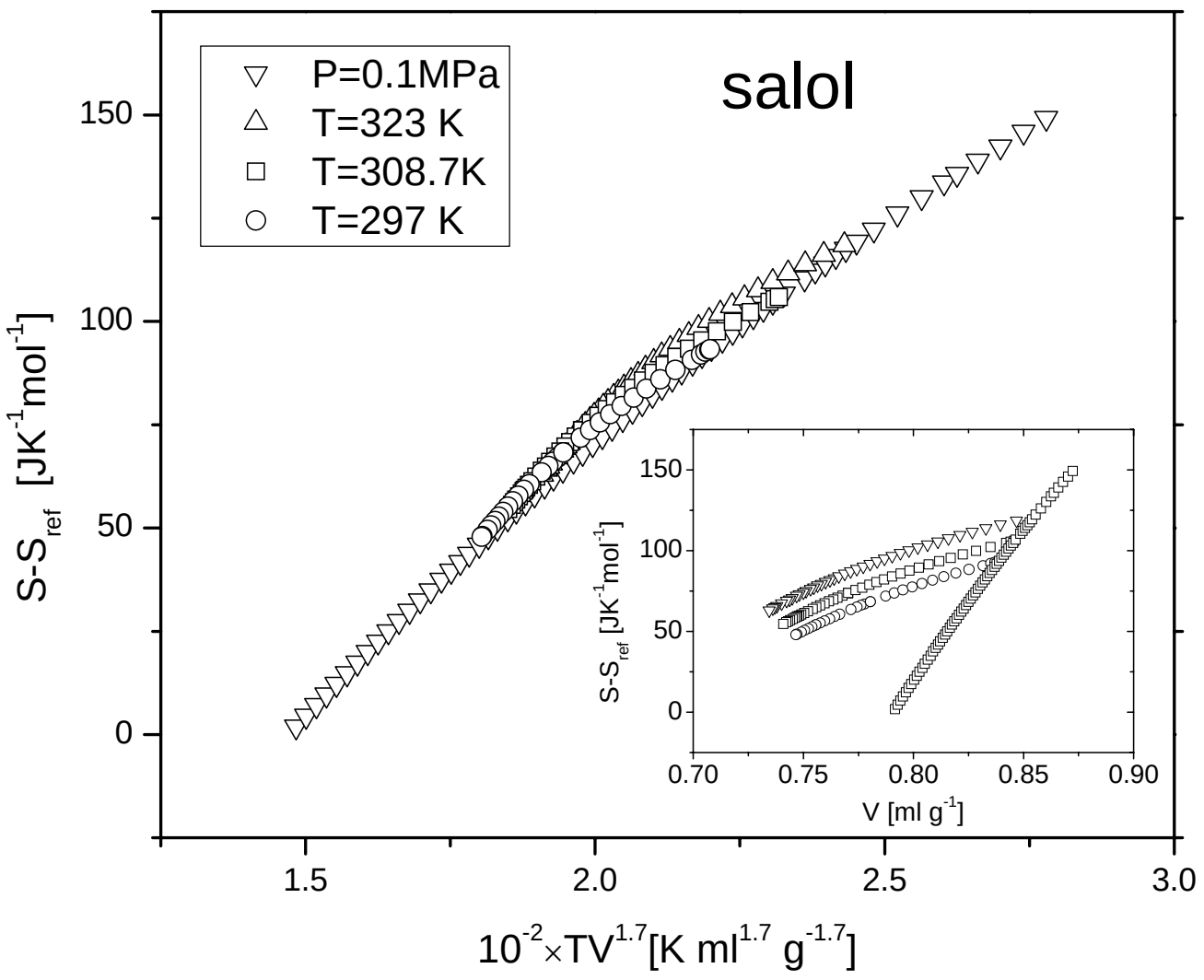

Figure 1

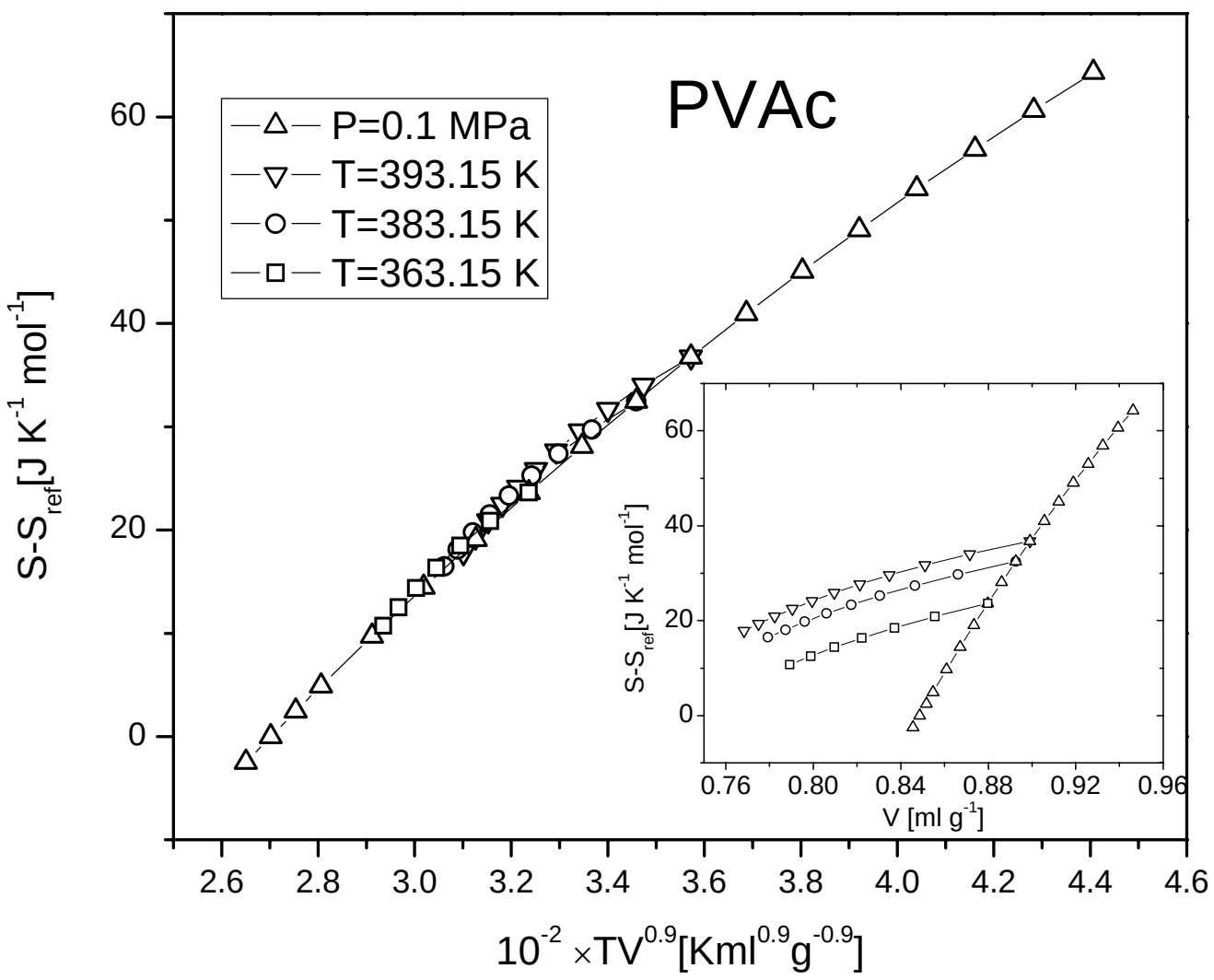

Figure 2

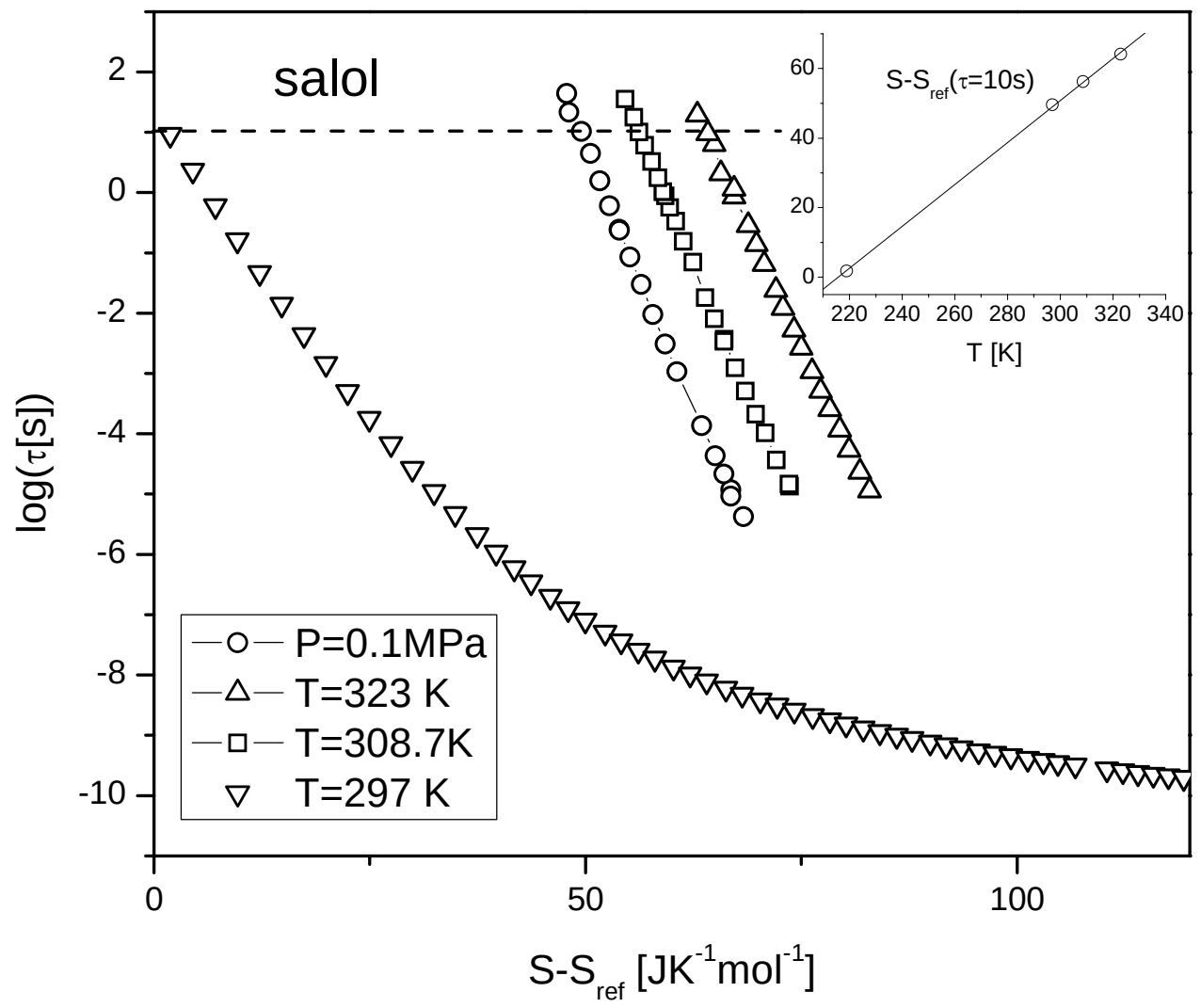

Figure 3

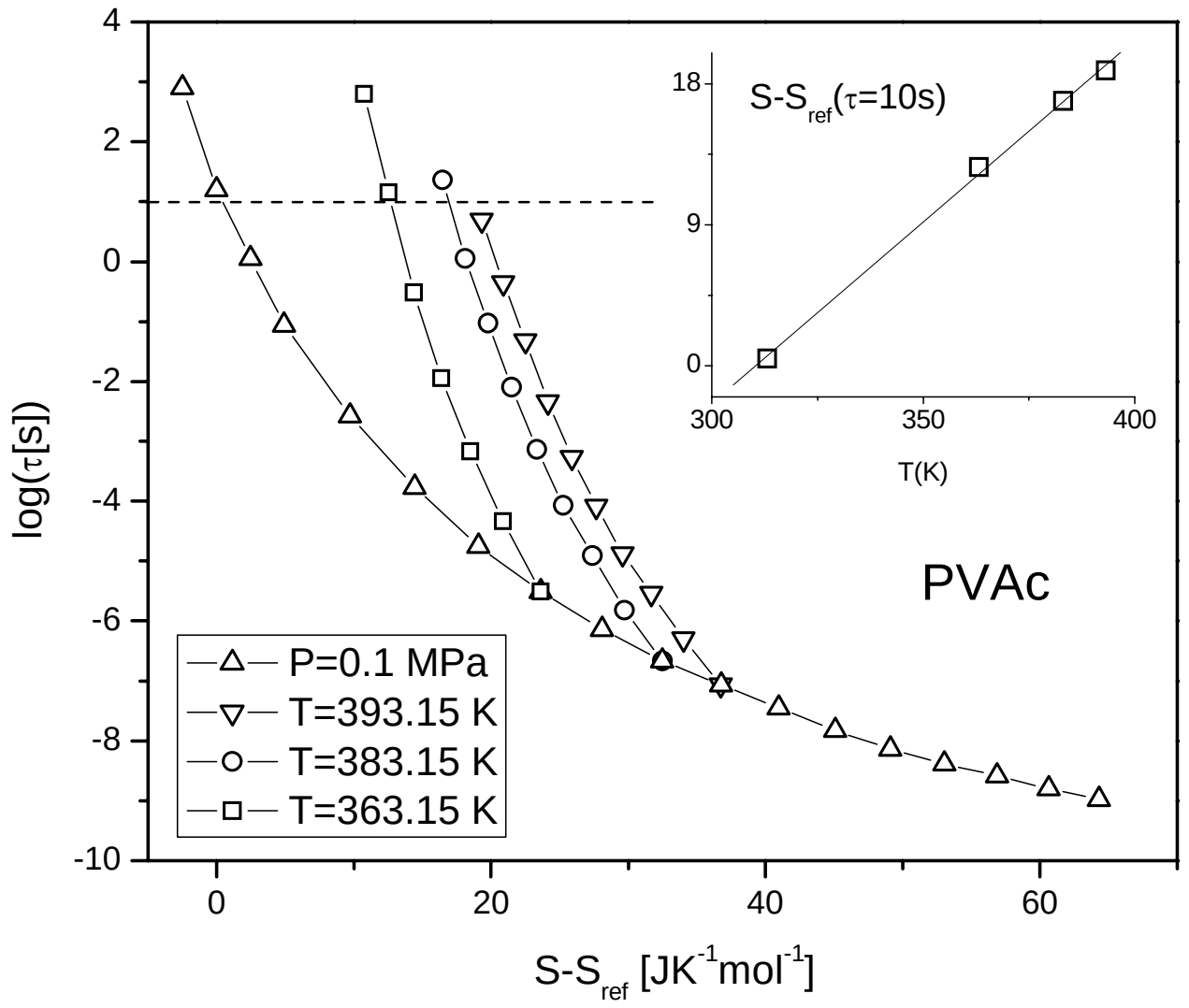

Figure 4

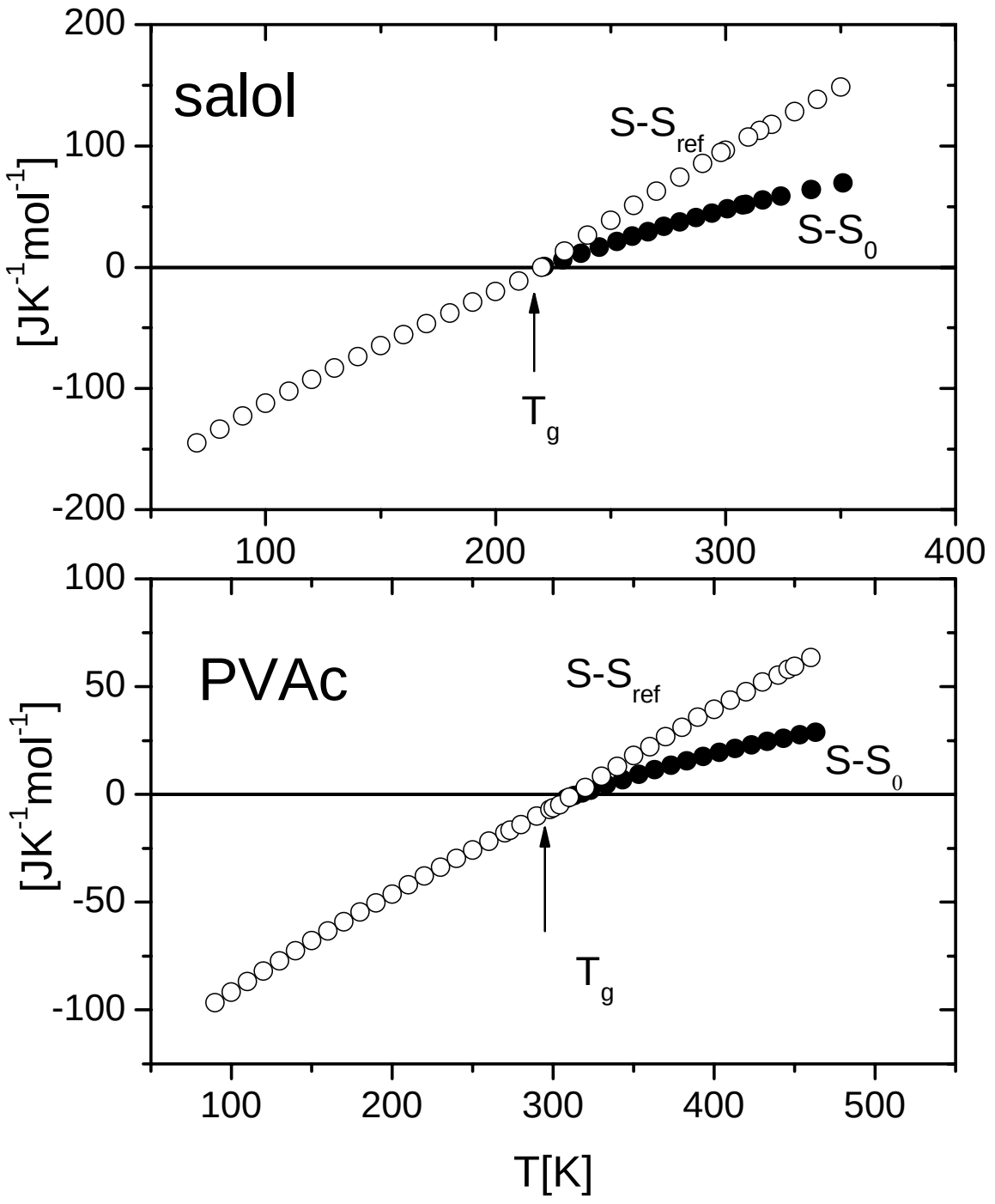

Figure 5

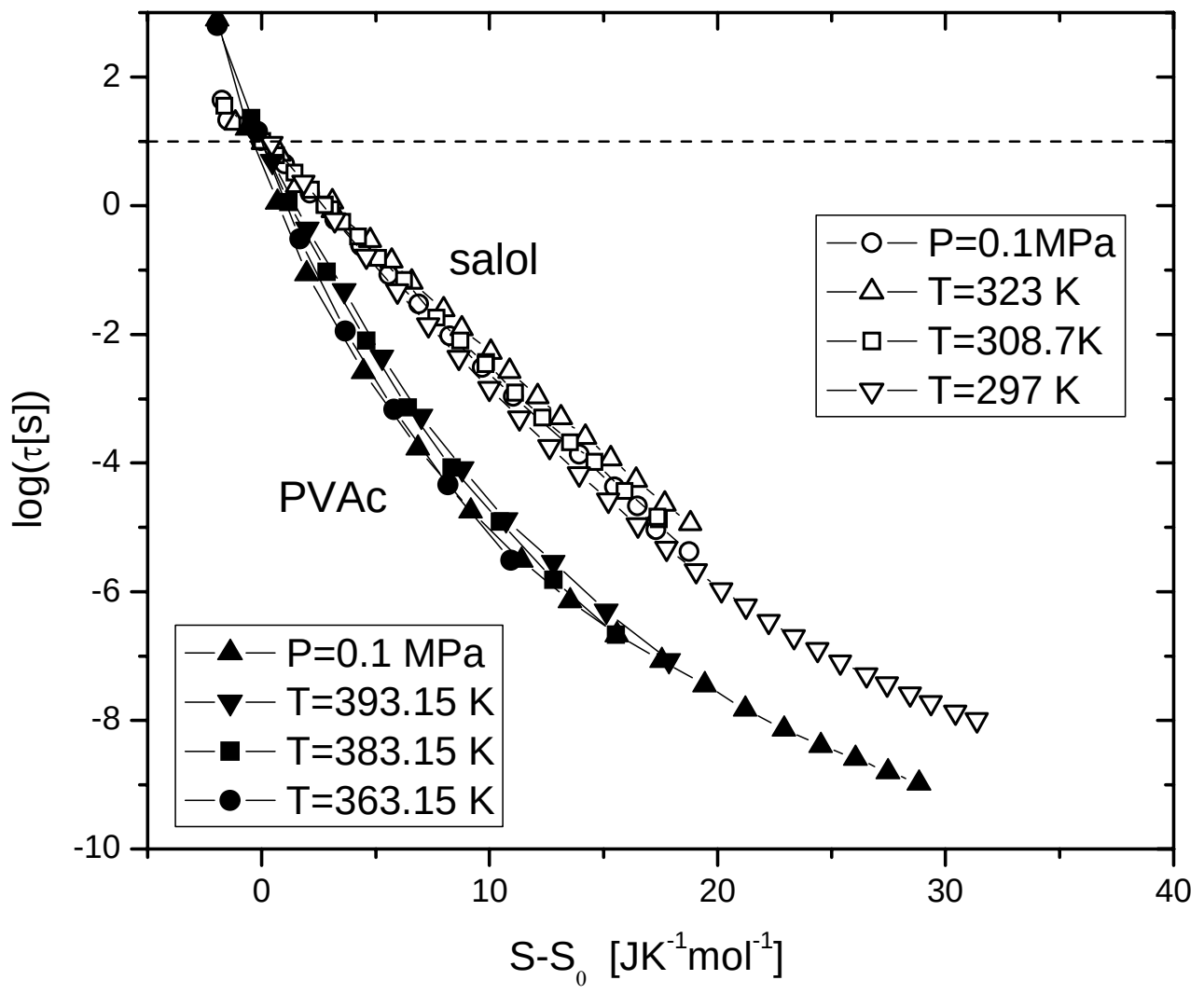

Figure 6

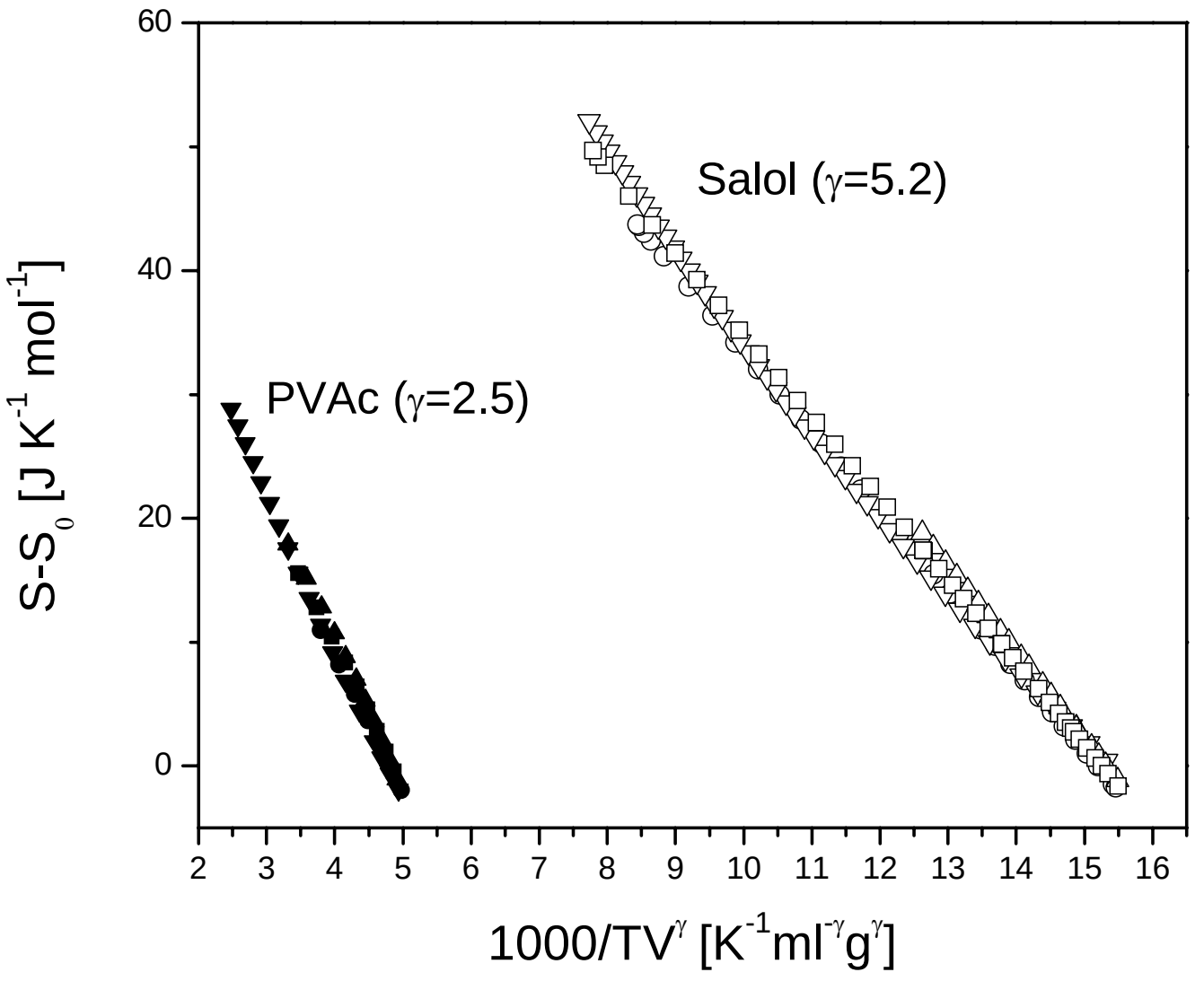

60
40
20
0
2 3 4 5 6 7 8 9 10 11 12 13 14 15 16
S-S0 [J K-1 mol-1]
1000/TVγ [K-1ml-γgγ]
Salol (γ=5.2)
PVAc (γ=2.5)

Figure 7